\documentclass[prl,reprint,twocolumn,showpacs,superscriptaddress]{revtex4-1}

\usepackage{bm,mathrsfs}
\usepackage{graphicx}
\usepackage{epsfig}
\usepackage{amsmath,bbm}
\usepackage{amsfonts,amssymb}
\usepackage{times}
\usepackage{dsfont}
\usepackage{enumitem}  
\usepackage{comment}
\usepackage[colorlinks,linkcolor=blue,citecolor=blue]{hyperref}
\newcommand{\eqn}[1]{
\begin{eqnarray}
	#1
\end{eqnarray}
}

\begin{document}
\title{Cavity Quantum Electrodynamics at Arbitrary Light-Matter Coupling Strengths} 
\author{Yuto Ashida}
\email{ashida@phys.s.u-tokyo.ac.jp}
\affiliation{Department of Physics, University of Tokyo, 7-3-1 Hongo, Bunkyo-ku, Tokyo 113-0033, Japan}
\affiliation{Institute for Physics of Intelligence, University of Tokyo, 7-3-1 Hongo, Tokyo 113-0033, Japan}
\affiliation{Department of Applied Physics, University of Tokyo, 7-3-1 Hongo, Bunkyo-ku, Tokyo 113-8656, Japan}
\author{Ata\c c $\dot{\mathrm{I}}$mamo$\breve{\mathrm{g}}$lu}
\affiliation{Institute of Quantum Electronics, ETH Zurich, CH-8093 Z{\"u}rich, Switzerland}
\author{Eugene Demler}
\affiliation{Department of Physics, Harvard University, Cambridge, MA 02138, USA}

\begin{abstract} 
Quantum light-matter systems at strong coupling are notoriously challenging to analyze due to the need to include states with many excitations in every coupled mode. We propose a nonperturbative approach to analyze light-matter correlations at all interaction strengths. The key element of our approach is a unitary transformation that achieves asymptotic decoupling of light and matter degrees of freedom in the limit where light-matter interaction becomes the dominant energy scale. In the transformed frame, truncation of the matter/photon Hilbert space is increasingly well-justified at larger coupling,  enabling one to systematically derive  low-energy effective models, such as tight-binding Hamiltonians. We demonstrate the versatility of our approach by applying it to concrete models relevant to electrons in crystal potential and electric dipoles interacting with a cavity mode. A generalization to the case of spatially varying electromagnetic modes is also discussed.
\end{abstract}

\maketitle

Understanding quantum systems with strong light-matter interaction has become a central problem in both fundamental physics and  quantum technologies \cite{CCT89}.
Recent experimental and theoretical advances in solid-state physics \cite{Basovaag1992,BJ19,HRJ04,KCS10,DCR10,CG13,OE15,CT16,JC17,AB17,KS18,RS18,GA18,PBGL19,KJ20,CT20,CE20,MNS20,TA192,RM14,SJ15,FeJ15,CA15,HD17,HD18,SMA18,SF19,CJB19,JDM19,RV19,MG19,MK19,WX19,YA20,LK20,CA20,DO20}, quantum optics \cite{RJM01,FP19,WA04,BA04,FDP10,MC14,HC17,BAn17,FDP17,YF17,YF172,LX18,SKR20,WK20,YT04,KG06,LG15,AR13,RD17,LEHUR2016808,KAF19,LBA20,SA20,BA04,CC05,LSD07,BJ09,CJ10,DLS14,PJS15,AS17,JT16,DB18,FS20,PP20,BM20}, and quantum chemistry \cite{ETW16,FJ17,TJR05,ST11,HJA12,CDM14,TA16,CR16,ZX17,SK18,MMLA19,Eiznereaax4482,C9SC04950A,Xiang665,MMLA18,TA19,RM18,GJ15,FJ15,FJ172} have made it possible to achieve strong coupling regimes in a variety of setups. In these systems, standard assumptions such as the rotating wave approximation can no longer be justified, and the inclusion of the diamagnetic $\hat{A}^2$ term or multilevel structure of matter becomes crucial.
Thus, quantized light and matter degrees of freedom must be treated on equal footing within the exact quantum electrodynamics (QED) Hamiltonian. Despite  considerable theoretical efforts,  a comprehensive formulation for analyzing such challenging problems at arbitrary coupling strengths is still lacking.

On another front, strongly correlated many-body systems have often been tackled by devising a unitary transformation that disentangles certain degrees of freedom, after which a simplified ansatz can be applied; a highly entangled quantum state in the original frame can then be expressed as a factorable state after the transformation. This general idea has been used in several contexts, such as analyzing quantum impurity systems \cite{LLP53,FH52,SR84,YA18L,*YA18B}, constructing low-energy effective models \cite{SJR66,GS93,WF94,BS11}, and solving many-body localization \cite{IJZ16} or electron-phonon problems \cite{TS20}.

The aim of this Letter is to extend this nonperturbative approach to strongly correlated light-matter systems, thus developing a consistent and versatile framework to seamlessly analyze arbitrary coupling regimes.  
Specifically, we propose to use a unitary transformation that asymptotically decouples light and matter in the strong-coupling limit. Our approach puts no limitations on the coupling strength and allows us to explore the full range of system parameters, including the regime where light-matter coupling dominates over all other relevant energy scales. Importantly, we construct a general way to systematically derive low-energy effective models by faithful level truncations, which remain valid at all coupling strengths. 
 This in particular provides a solution to the long-standing controversy  \cite{LWE52,RK75,Keeling_2007,NP10,CL12,VA14,GMF17,BSJ17,DB182,AGM19,AGM20,SA192,*SA193,SA19,DSO19,LJ20,SC20,TMAD20} about which frame is best suited for studying strong-coupling physics.
We demonstrate the versatility of our formalism by applying it to specific models relevant to materials and atomic systems in cavity QED.

\begin{figure}[b]
\includegraphics[width=70mm]{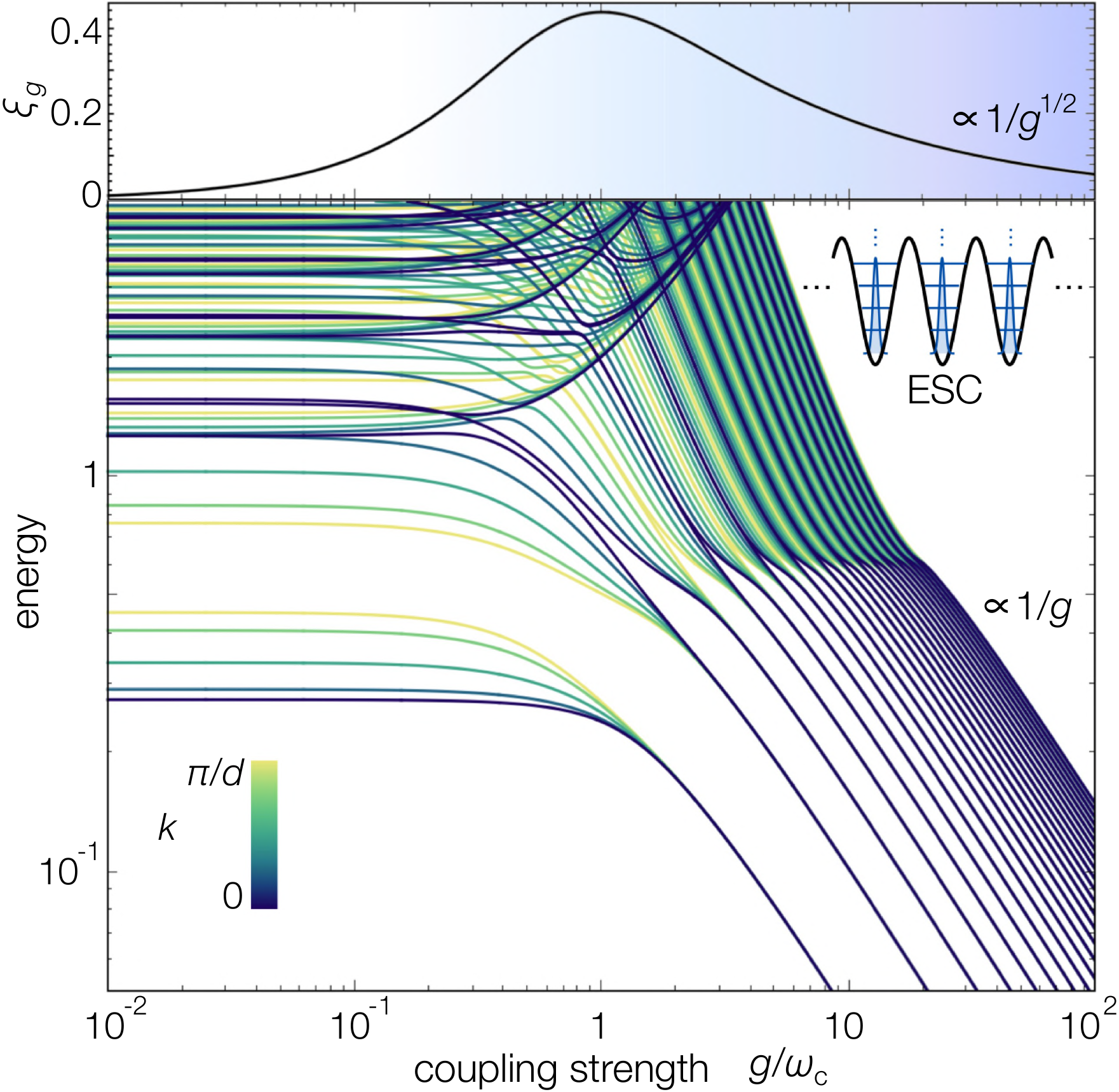} 
\caption{\label{fig1}
(Top) Effective parameter $\xi_g$ characterizing interaction strength in the asymptotically decoupled frame against the bare light-matter coupling  $g$. (Bottom) Exact spectrum obtained by diagonalizing Eq.~\eqref{HU}, or equivalently \eqref{HUs}, for an electron in periodic potential. In the extremely strong coupling (ESC) regime, it exhibits equally spaced flat bands narrowing as $\propto\!1/g$, corresponding to localized electrons with the large renormalized mass (inset). Numerical values are shown in the unit $\omega_c\!=\!\hbar\!=\!m\!=\!1$ throughout this Letter. The potential depth and lattice constant are $v\!=\!5$ and $d\!=\!4$, respectively.  
}
\end{figure}

{\it Asymptotic decoupling of light-matter interaction.---} 
To illustrate the main idea, we first focus on a one-dimensional many-body system coupled to a single electromagnetic mode; a generalization to higher-dimensional systems with spatially varying electromagnetic modes will be given later. 
We start from the QED Hamiltonian in the Coulomb gauge:
\eqn{\label{HC}
\hat{H}_{{\rm C}}\!=\!\int \!dx\,\hat{\psi}_{x}^{\dagger}\left[\frac{(-i\hbar\partial_{x}\!-\!q\hat{A})^{2}}{2m}\!+\!V(x)\right]\hat{\psi}_{x}\!+\!\hbar\omega_{c}\hat{a}^{\dagger}\hat{a}\!+\!\hat{H}_{||},\nonumber\\
}
where $\hat{\psi}_{x}$ ($\hat{\psi}_{x}^\dagger$) is the annihilation (creation) operator of fermions of mass $m$ and charge $q$ at position $x$, and $V(x)$ is an arbitrary external potential. 
Equation~\eqref{HC} describes the coupling between electrons and 
 a cavity electromagnetic mode with  frequency $\omega_c$ and the vector potential operator $\hat{A}={\cal A}(\hat{a}+\hat{a}^{\dagger})$, where ${\cal A}$ is the mode amplitude and $\hat{a}$ ($\hat{a}^\dagger$) is the annihilation (creation) operator of photons. The instantaneous Coulomb interaction is given by $\hat{H}_{||}=\int dxdx'\,q^{2}\hat{\psi}_{x}^{\dagger}\hat{\psi}_{x'}^{\dagger}\hat{\psi}_{x'}\hat{\psi}_{x}/4\pi\epsilon_{0}|x-x'|$. We rewrite $\hat{H}_{\rm C}$ as  
\eqn{
\hat{H}_{{\rm C}}&=&\int dx\,\hat{\psi}_{x}^{\dagger}\left[-\frac{\hbar^{2}\partial_{x}^{2}}{2m}+V(x)\right]\hat{\psi}_{x}+\hbar\Omega\hat{b}^{\dagger}\hat{b}+\hat{H}_{||}\nonumber\\
&{}&-gx_{\Omega}\int dx\,\hat{\psi}_{x}^{\dagger}(-i\hbar\partial_{x})\hat{\psi}_{x}\,(\hat{b}+\hat{b}^{\dagger}),
}
where $\Omega=\sqrt{\omega_{c}^{2}+2Ng^{2}}$ is the dressed photon frequency with the particle number $N$ \footnote{We note that, in solid-state systems, the electron density $N/V$ with $V$ being the volume should be a natural quantity in the dressed frequency, where the volume factor arises from the mode amplitude ${\cal A}\propto1/\sqrt{V}$.} and the coupling strength $g=q{\cal A}\sqrt{{\omega_{c}}/{m\hbar}}$, and $x_{\Omega}=\sqrt{\hbar/m\Omega}$ is a characteristic length relevant both in weak and strong coupling regimes.
Here, the photon part has been diagonalized by a Bogoliubov transformation: $\hat{b}+\hat{b}^{\dagger}=\sqrt{\Omega/\omega_{c}}\,(\hat{a}+\hat{a}^{\dagger})$ \footnote{The coefficients in the Bogoliubov transformation need to be real-valued in order to diagonalize the quadratic photon Hamiltonian.}. 

To asymptotically decouple light and matter degrees of freedom, we propose to use a unitary transformation 
\eqn{\label{unitary}
\hat{U}=\exp\left[-i\xi_{g}\int_{-\infty}^\infty dx\,\hat{\psi}_{x}^{\dagger}(-i\partial_{x})\hat{\psi}_{x}\,\hat{\pi}\right],
}
where $\xi_{g}=gx_{\Omega}/\Omega$ is the effective length scale characterized by the coupling strength $g$ and $\hat{\pi}=i(\hat{b}^{\dagger}-\hat{b})$. 
The transformation~\eqref{unitary} is reminiscent of the Lee-Low-Pines transformation used for polaronic systems \cite{LLP53}, and leads to the Hamiltonian $\hat{H}_{U}\equiv\hat{U}^{\dagger}\hat{H}_{{\rm C}}\hat{U}$ given by
\eqn{\label{HU}
\hat{H}_{U}\!&=&\!\!\int dx\,\hat{\psi}_{x}^{\dagger}\!\left[-\frac{\hbar^{2}\partial_{x}^{2}}{2m}\!+\!V\left(x+\xi_{g}\hat{\pi}\right)\!\right]\hat{\psi}_{x}\!+\!\hbar\Omega\hat{b}^{\dagger}\hat{b}\!+\!\hat{H}_{\parallel}\nonumber\\
&&-\frac{\hbar^{2}g^{2}}{m\Omega^{2}}\left[\int dx\,\hat{\psi}_{x}^{\dagger}(-i\partial_{x})\hat{\psi}_{x}\right]^2,
}
where the light-matter interaction is now absorbed by the potential term as the  shift $\xi_{g}\hat{\pi}$ of the electron coordinates. 
  Physically, the unitary operator~\eqref{unitary} changes a reference frame in such a way that quantum particles no longer interact with the electromagnetic mode through the usual minimal coupling, but through the gauge-field dependent shift of the electron coordinates and the associated quantum fluctuations in the external potential. 
Thus, in the transformed frame the effective strength of the light-matter interaction is characterized by $\xi_g$ instead of the original coupling  $g$. 
Remarkably, as shown in the top panel of Fig.~\ref{fig1}, $\xi_g$ remains small  over the entire region of $g$ and, in particular, vanishes as $\xi_g\!\propto\! g^{-1/2}$ in the strong-coupling limit $g\!\to\!\infty$. 
For this reason, we shall call the present frame as the {\it asymptotically decoupled} (AD) frame; the identification of the AD Hamiltonian~\eqref{HU} is the first main result of this Letter. 

Several remarks are in order. First, a specific form of $\xi_g$ can depend on polarization of an electromagnetic mode. For instance, when matter is coupled to a circularly polarized mode, $\xi_g$ vanishes as $\xi_g\!\propto\! g^{-1}$ provided that the coupling $g$ is sufficiently large  \cite{SM1}.  
Second, we note that the transformation~\eqref{unitary} preserves the translational symmetry of the (bare) matter Hamiltonian. This should be compared to, e.g., the Power-Zienau-Woolley (PZW) frame \cite{PZ59,WRG71} in which such symmetry is broken due to the additional terms in the transformed Hamiltonian \cite{SM1,DB182}. Third, in view of our definition of the coupling strength $g$, the so-called ultrastrong (deep strong) coupling regime should approximately correspond to $g\gtrsim 0.3$ ($g\gtrsim 3$). 
 Below we show that further increase of $g$ leads to the new regime, namely, the {\it extremely strong coupling} (ESC) regime. In the latter, truncation of matter/photon levels can no longer be justified in the conventional frames, but is asymptotically exact in the AD frame as discussed in detail below.

{\it General properties at extremely strong coupling.---} 
From now on, we focus on the single-electron problems and delineate universal spectral features in the ESC regime; the role of electron interactions will be discussed in a future publication. The AD-frame Hamiltonian is then simplified to
\eqn{\label{HUs}
\hat{H}_{U}=\frac{\hat{p}^{2}}{2m_{{\rm eff}}}+V\left(x+\xi_{g}\hat{\pi}\right)+\hbar\Omega\hat{b}^{\dagger}\hat{b},
}
where renormalization of the effective mass $m_{{\rm eff}}=m[1+2(g/\omega_{c})^{2}]$ exactly arises from the last term in Eq.~\eqref{HU}; this renormalization becomes even more prominent in a many-body case \cite{SM1}. Note that the last term in Eq.~\eqref{HU} also generates the interaction term $\propto\hat{\psi}^\dagger\hat{\psi}^\dagger\hat{\psi}\hat{\psi}$, which, however, does not affect single-electron systems considered below. 

One can understand the key spectral features of $\hat{H}_U$ in the ESC regime as follows. In the limit of large $g$, the renormalized photon frequency $\Omega$ becomes large, while the effective light-matter coupling, characterized by $\xi_g$, eventually decreases. Thus, in the strong-coupling limit, the lowest-energy eigenstates $|\Psi_U\rangle$ of $\hat{H}_U$ are well approximated by a product state: 
\eqn{
|\Psi_U\rangle\simeq|\psi_U\rangle\,|0\rangle_\Omega,
}
where $|\psi_U\rangle$ is an eigenstate of $\hat{p}^2/2m_{\rm eff}\!+\!V(x)$, and $|0\rangle_{\Omega}$ is  the dressed-photon vacuum.
Now, suppose that potential $V$ has well-defined local minima, around which it can be expanded as  $\delta V\propto x^2$. Since the effective mass rapidly increases as $m_{\rm eff}\propto g^2$,   $|\psi_U\rangle$ is tightly localized around the potential minima. The low-lying spectrum of $\hat{H}_U$ thus reduces to that of the harmonic oscillator with narrowing level spacing $\delta E\propto 1/g$.

The above argument shows that, in the AD frame, an energy eigenstate can be well approximated by a product of light and matter states.  Nevertheless, they are still strongly entangled in the original frame. To see this, we consider an eigenstate $|\Psi_{\rm C}\rangle=\hat{U}|\Psi_U\rangle$ of the Coulomb-gauge Hamiltonian $\hat{H}_{\rm C}$:
\eqn{
\label{eigC}
|\Psi_{\rm C}\rangle=\hat{U}\int dp\,\psi_{p}|p\rangle|0\rangle_{\Omega}=\int dp\,\psi_{p}|p\rangle\hat{D}_{\xi_{g}p}|0\rangle_{\Omega},}
where $\int dp\, \psi_p|p\rangle\!=\!|\psi_U\rangle$ is the AD-frame eigenstate expressed in the momentum basis,  and $\hat{D}_\beta\!=\!e^{\beta\hat{b}^{\dagger}-\beta^{*}\hat{b}}$ is the displacement operator. In the ESC regime,  $|\psi_U\rangle$ has vanishingly small  
variance $\sigma_x\!\propto\! 1/g$; accordingly, the momentum distribution $|\psi_p|^2$ is very broad with variance $\sigma_p\!\propto\! g$,
showing that the Coulomb-gauge eigenstate~\eqref{eigC} is a highly entangled state consisting of superposition of coherent states with large photon occupancy determined by the particle momentum.

{\it Difficulties of level truncations in conventional frames.---} 
The AD frame readily allows us to elucidate the origin of difficulties for level truncations in the Coulomb gauge~\cite{DB182,DSO19,LJ20,TMAD20}.   
Namely, if we expand a tightly localized state $|\psi_U\rangle$ in terms of eigenstates of $\hat{p}^2/2m+V$ with the {\it bare} mass $m$, we will find substantial contribution from high-energy electron states. 
This point can be seen from Eq.~\eqref{eigC}, which contains large-momentum eigenstates. Thus, any analysis performed in the Coulomb gauge, which uses a fixed UV cutoff for electron states, should become invalid at sufficiently strong coupling.  
While the use of the PZW frame can partially mitigate the limitations  \cite{CA15,LJ20,DO20} and can be valid up to ultrastrong/deep strong coupling regimes \cite{DB182,SA192}, it is ultimately constrained by the same restrictions,  especially in the ESC regime. This holds true even when high-lying states appear to be reasonably out of resonance. Moreover, both the mean and fluctuation of the photon number in the PZW frame increases as $\overline{n},\delta n\!\propto\!g$. Thus, the number of photon states required to diagonalize the Hamiltonian diverges at large $g$, making photon-level truncation (that is unavoidable in actual calculations) ill-justified in the deep- or extremely-strong coupling regimes. 
Altogether, as long as one relies on the conventional frames, we conclude that  effective models derived by level truncations, such as tight-binding models or the quantum Rabi model, must inevitably break down when $g$ becomes sufficiently large. 

In contrast, the AD frame~\eqref{HUs} introduced here provides a simple solution to this problem. Specifically, matter-level truncation, i.e., tight-binding approximation, is increasingly well-justified in $\hat{H}_U$ at larger $g$, owing to tighter localization of the wavefunction $|\psi_U\rangle$.
Similarly, due to the photon dressing and asymptotic decoupling, one can always truncate high-lying photon levels; indeed, the mean photon number remains very small over the entire region of $g$ and, in particular, vanishes in the ESC limit.  Below we demonstrate such versatility of the AD frame by applying it to concrete models relevant to quantum electrodynamical materials and atomic dipoles.

\begin{figure}[t]
\includegraphics[width=86mm]{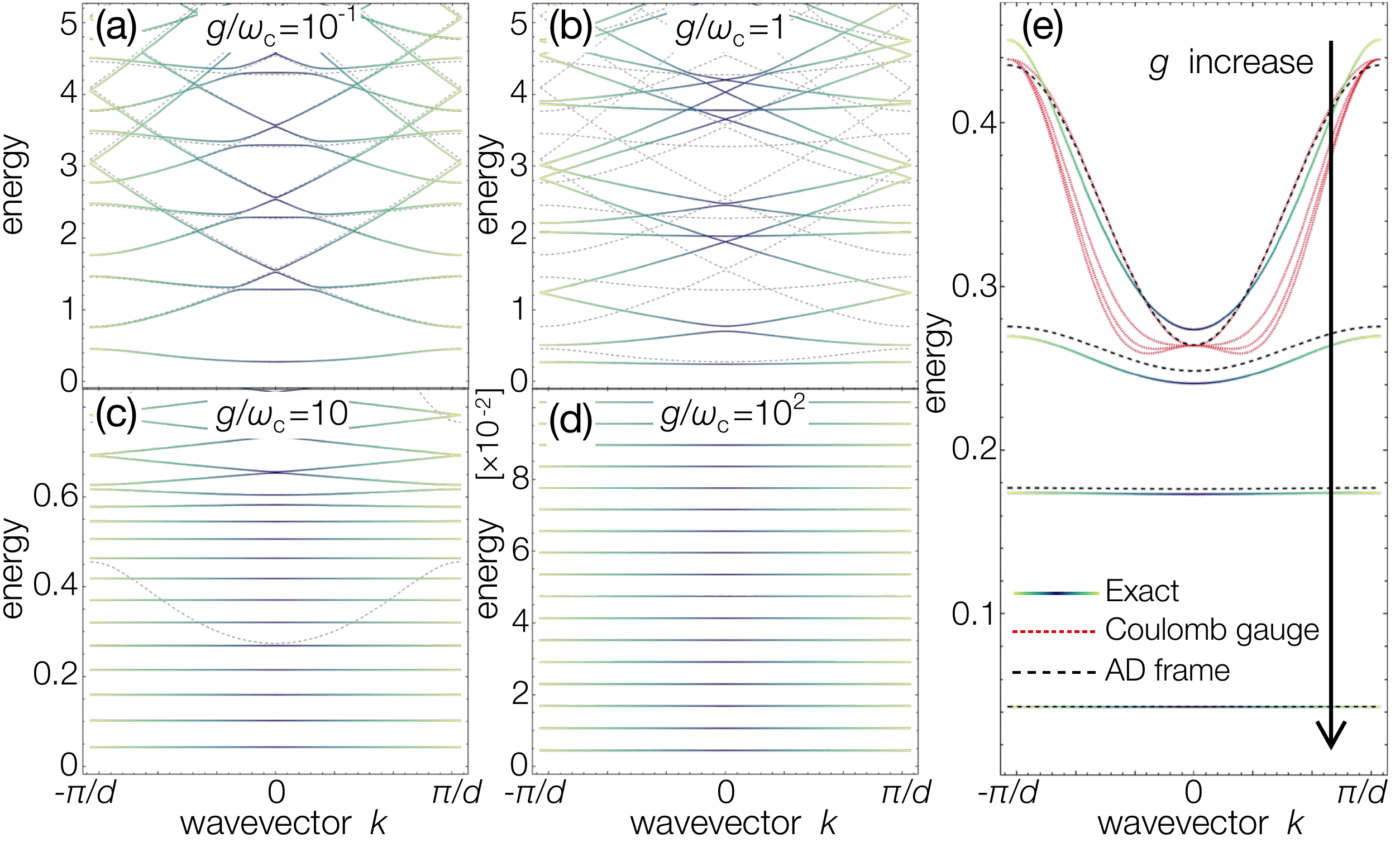} 
\caption{\label{fig2}
(a-d) Exact electron-polariton bands obtained by diagonalizing Eq.~\eqref{HUs}. Gray dashed curves indicate dispersions at $g\!=\!0$. (e)~Comparisons between the exact results and the tight-binding models at $g\!=\!0.1,1,2,10$ from top to bottom. Black dashed (red dotted) curves indicate the tight-binding results in the asymptotically decoupled frame (Coulomb gauge). We set $v\!=\!5$ and $d\!=\!4$ in (a-e).
}
\end{figure}

{\it Application to solid-state systems.---} 
We first consider an electron in periodic potential and discuss the formation of electron-polariton band structures. To be concrete, we assume $V=v[1+\cos\left(2\pi x/d\right)]$ with $d$ and $v$ being the lattice constant and potential depth, respectively. Since the AD frame preserves the translational symmetry, Bloch's theorem remains valid and every eigenvalue of $\hat{H}_U$ has a well-defined crystal wavevector $k\in[-\pi/d,\pi/d)$.      
Figures~\ref{fig1} and \ref{fig2}a-d show the obtained {\it exact} eigenspectra  at different coupling strengths $g$, in the sense that matter/photon-energy cutoffs are taken to be large enough such that the results  are converged.  As $g$ is increased, the bands become increasingly flat and form equally spaced spectra with energy spacing narrowing $\propto\!1/g$, which is fully in accord with the universal spectral features discussed earlier. While the signature of band flattening can emerge already at deep strong coupling \cite{LJ20}, the drastic level narrowing/softening of the whole excitation spectrum is one of the key distinctive features of the ESC regime [see Fig.~\ref{fig1}]. 

To construct the effective low-energy Hamiltonian, we derive the tight-binding model by projecting the continuum system on the lowest-band Wannier orbitals. Specifically, we first expand a matter state in terms of the Wannier basis,
\eqn{
\hat{\psi}_{x}=\sum_{j}w_{j}(x)\hat{c}_{j},
}  
where $w_j$ is the Wannier function at site $j$ for the lowest band of $\hat{p}^2/2m_{\rm eff}+V$ with the {\it effective} mass, and $\hat{c}_j$ is the corresponding annihilation operator. We then consider a manifold spanned by product states of these Wannier orbitals and an arbitrary photon state. Projecting $\hat{H}_U$ on this manifold and considering the leading contributions, we obtain the tight-binding Hamiltonian in the AD frame as \cite{SM1}
\eqn{
\hat{H}_{U}^{{\rm TB}}&=&(t_{g}+t'_{g}\hat{\delta}_g)\sum_{i}(\hat{c}_{i}^{\dagger}\hat{c}_{i+1}+{\rm h.c.})\nonumber\\
&&+(\mu_{g}+\mu'_{g}\hat{\delta}_g)\sum_{i}\hat{c}_{i}^{\dagger}\hat{c}_{i}+\hbar\Omega\hat{b}^{\dagger}\hat{b},
} 
where $t_{g}\!=\!\int\frac{dk}{K}\epsilon_{k,g}e^{ikd}$ is the effective hopping parameter with $\epsilon_{k,g}$ being the lowest-band energy of  $\hat{p}^2/2m_{\rm eff}\!+\!V$ and $K\!=\!2\pi/d$, and $\mu_{g}\!=\!\int\frac{dk}{K}\epsilon_{k,g}$ is the effective chemical potential. The electromagnetically induced fluctuation of potential causes the terms with $\hat{\delta}_g\!=\!\cos\left(K\xi_{g}\hat{\pi}\right)\!-\!1$, $t'_{g}\!=\!v\int dx\,w_{i}^{*}\cos(Kx)w_{i+1}$, and $\mu'_{g}\!=\!v\int dx\,w_{i}^{*}\cos(Kx)w_{i}$.

Figure~\ref{fig2}e shows that this surprisingly simple tight-binding model (black dashed curve) accurately predicts the exact spectrum (solid curve) at any $g$, and asymptotically becomes exact in the strong-coupling limit as expected. For the sake of comparison, we show the tight-binding results in the Coulomb gauge (red dotted curve), which are obtained by projecting $\hat{H}_{\rm C}$ onto the lowest band of $\hat{p}^2/2m+V$ with the $bare$ mass \cite{SM1}. While this na\"{\i}ve tight-binding model is valid when $g\!\lesssim\!0.1$, it completely misses key features  at larger $g$, such as band flattening and level narrowing in the whole excitation spectrum. Physically, this drastic failure originates from ill-justified truncation of strongly entangled high-lying light-matter states in the original frame [cf. Eq.~\eqref{eigC}].   

These results clearly demonstrate that a choice of the frame is essential  to construct an accurate tight-binding model in strong-coupling regimes. The AD frame solves this issue by performing the projection {\it after} the unitary transformation, which effectively realizes suitable nonlinear truncation in the Coulomb gauge.  The most general form of the AD-frame tight-binding Hamiltonian is given by
\eqn{\label{TBU}
\hat{H}_{U}^{{\rm TB}}\!\!=\!\!\sum_{ij\nu\lambda}\!t_{ij\nu\lambda}\hat{c}_{i\nu}^{\dagger}\hat{c}_{j\lambda}\!+\!\!\!\sum_{lij\nu\lambda}\!t_{ij\nu\lambda}'^{(l)}\hat{c}_{i\nu}^{\dagger}\hat{c}_{j\lambda}\hat{\pi}^{l}\!\!+\!\hbar\Omega\hat{b}^{\dagger}\hat{b},
}  
where $\nu$ $(\lambda)$ labels internal degrees of freedom in each unit cell $i$ $(j)$, and  
$t_{ij\nu\lambda}\!\!=\!\!\int dx w_{i\nu}^{*}[{\hat{p}^{2}}/{2m_{{\rm eff}}}\!+\!V]w_{j\lambda}$, $t_{ij\nu\lambda}'^{(l)}\!\!=\!\!\frac{\xi_{g}^{l}}{l!}\int dxw_{i\nu}^{*}V^{(l)}w_{j\lambda}$ with $w_{i\nu}$ being the corresponding Wannier functions, and $V^{(l)}$ is the $l$-th derivative of  $V$ with $l\!=\!1,2,\ldots$ The renormalized parameters $t,t'^{(l)}$ nonperturbatively depend on $g$ through the nonlinear truncation. Higher-order terms at larger $l$ contribute less to eigenspectrum owing to smallness of $\xi_g$ [cf. Fig.~\ref{fig1}], which enables a systematic approximation when necessary. The minimal tight-binding Hamiltonian~\eqref{TBU}, which should be valid at arbitrary coupling strengths and even under disorder,  provides the material counterpart of the quantum Rabi model. Its derivation is the second main result of this Letter. 

For a general periodic potential, the calculation of matrix elements of the shifted potential $V(x\!+\!\xi_g\hat{\pi})$ can be separated into light and matter parts, after which the standard procedures can be used \cite{SM1}. Even when a potential is not translationally invariant, one can expand it as $V(x\!+\!\xi_g\hat{\pi})\simeq V(x)\!+\!\sum_{l=1}^{l_{\rm max}}V^{(l)}(x)\xi_g^l\hat{\pi}^l$. The truncation order $l_{\rm max}$ should scale inversely with $g$ due to decreasing $\xi_g$. This expansion should be valid unless a potential has singular spatial dependence and expansion coefficients are not systematically suppressed at higher orders.

{\it Application to atomic dipoles.---} 
\begin{figure}[t]
\includegraphics[width=86mm]{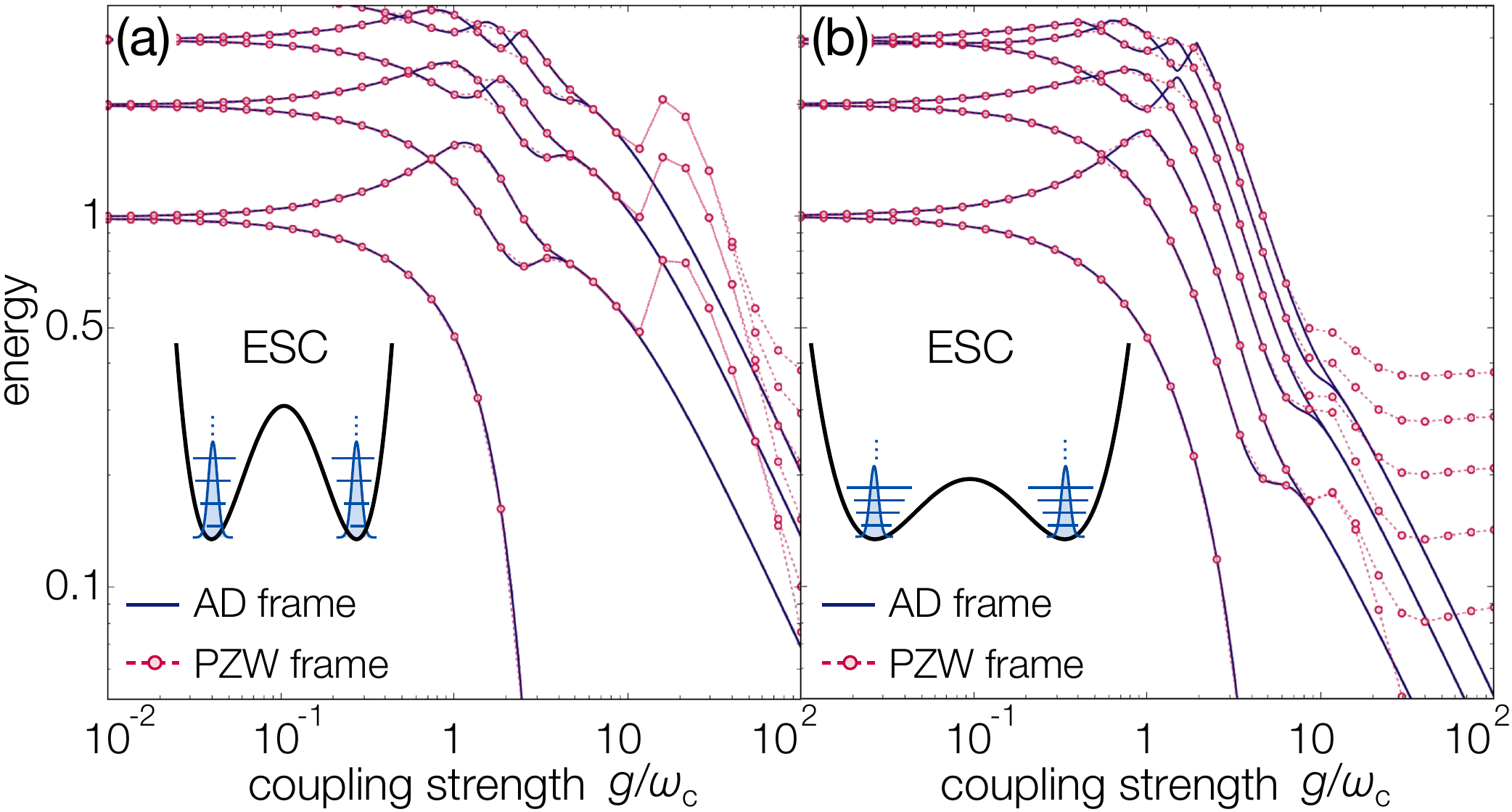} 
\caption{\label{fig3}
Low-energy spectra for (a) deep and (b) shallow double-well potentials with photon-number cutoff $n_c\!=\!100$. Blue solid curves (red dotted curves) show the results in the AD frame (PZW frame). We choose the parameters (a) $\lambda\!=\!50$, $\mu\!=\!95$ and (b) $\lambda\!=\!3$, $\mu\!=\!3.85$ such that the transition frequency of the two lowest matter levels is resonant with $\omega_c$ in each case.   
}
\end{figure}
We next apply the AD frame to the case of a quantum particle in double-well potential $V\!=\!-\lambda x^2/2\!+\!\mu x^4/4$, which is a standard model for the electrical dipole moment. Blue solid curves in Fig.~\ref{fig3}a,b show the  spectra obtained in the AD frame at different potential depths, where the results efficiently converge already at a low photon-number cutoff $n_c\!\sim\! 5$-$10$ \cite{SM1}.  The spectra at ESC exhibit the universal features discussed above, i.e., energies become doubly degenerate corresponding to two wells and are equally spaced with narrowing $\propto\! 1/g$ due to tight localization around the minima [cf. insets]. 

As discussed earlier, truncation of high-lying photon states should eventually be invalid in conventional frames. We demonstrate this by comparing to results obtained in the PZW frame, $\hat{H}_{{\rm PZW}}\!=\!\hat{U}_{{\rm PZW}}^{\dagger}\hat{H}_{{\rm C}}\hat{U}_{{\rm PZW}}$ with $\hat{U}_{\rm PZW}\!=\!\exp(iqx\hat{A}/\hbar)$,  
 at a large cutoff $n_c\!=\!100$ (red dotted curves in Fig.~\ref{fig3}). Notably, the PZW frame dramatically fails in the ESC regime, which has its root in the rapid increase of mean-photon number due to strong light-matter entanglement 
and sizable probability amplitudes of high photon-number states  \cite{SM1}. We remark that matter-level cutoff is taken to be sufficiently large such that the results converge because the strong light-matter entanglement also invalidates matter-level truncation in the Rabi-type descriptions.  Since any actual calculation must resort to finite  cutoffs, these results indicate the fundamental difficulties of the conventional frames in the ESC regime.

{\it Beyond the single-mode description.---} 
While the single-mode description can be justified in, e.g., an LC-circuit resonator \cite{YF172}, it may fail when more than one cavity mode must be included depending on the cavity geometry. The unitary transformation~\eqref{unitary} can be generalized to such a case with spatially varying electromagnetic modes:
\eqn{\hat{U}=\exp\left[-i\frac{\hat{\boldsymbol{p}}}{\hbar}\cdot\sum_{\alpha}\boldsymbol{\xi}_{\alpha}\hat{\pi}_{\alpha}(\boldsymbol{x})\right],}
where $\alpha$ labels multiple modes and the electromagnetic fields now depend on position $\boldsymbol{x}$. 
At the leading order, the transformed Hamiltonian is 
\eqn{\label{HUmulti}
\hat{H}_{U}&=&\frac{\hat{\boldsymbol{p}}^{2}}{2m}-\sum_{\alpha}\frac{\left(\hat{\boldsymbol{p}}\cdot\boldsymbol{\zeta}_{\alpha}\right)^{2}}{\hbar\Omega_{\alpha}}+V\bigl(\boldsymbol{x}+\sum_{\alpha}\boldsymbol{\xi}_{\alpha}\hat{\pi}_{\alpha}\left(\boldsymbol{x}\right)\bigr)\nonumber\\
&&+\sum_{\alpha}\hbar\Omega_{\alpha}\hat{b}_{\alpha}^{\dagger}(\boldsymbol{x})\hat{b}_{\alpha}(\boldsymbol{x}),
}
where $\boldsymbol{\zeta}_\alpha$ is the effective polarization vector of mode $\alpha$ \cite{SM1}. 
This simple expression is valid when field variation is small compared to the effective length scale, i.e., $k|\boldsymbol{\xi}|\!\ll\! 1$ with $|\nabla\hat{b}|\!\sim \!k\hat{b}$; this condition is independent of system size and much less restrictive than, e.g., the dipole approximation, owing to smallness of $\xi_g$.  We remark that light-matter decoupling due to the inhomogeneous diamagnetic term has previously been studied in the case of the quadratic Hamiltonian \cite{DLS14}.

{\it Discussions.---}  In the limit of classical electromagnetic fields, our  transformation~\eqref{unitary} can be compared with the Kramers-Henneberger (KH) transformation, which was used to analyze atoms subject to intense laser fields \cite{KHA56,HWC68}. Besides the full quantum treatment given here, one important difference is that the KH transformation does not take into account the diamagnetic $A^2$ term other than its contribution to ponderomotive forces appearing in spatially inhomogeneous laser profiles.
 In our quantum setting, the asymptotic light-matter decoupling emerges only after the diamagnetic term is consistently included through the Bogoliubov transformation. 

With the advent of new materials and subwavelength cavity designs, it is now possible to explore ultra/deep strong coupling regimes of light-matter interaction and possibilities for further extending the interaction strength. We expect our results to be applicable in the analysis of mono- or (twisted) bilayer-2D materials embedded in high quality-factor lumped-element terahertz cavities \cite{KJ20}, where a single mode of the electromagnetic field is isolated from higher-energy Fabry-Perot-like confined modes, as well as the electromagnetic continuum. Signatures of the level narrowing/softening in the ESC regime are already appreciable  around $g/\omega_c\gtrsim 5$, which can be realized with current experimental techniques \cite{AB17,YF172,MNS20}.

In summary, we presented a new formulation~\eqref{HU} of strongly correlated light-matter systems that is applicable to both quantum electrodynamical materials and atomic systems. Since this is a nonperturbative approach, it is valid at arbitrary coupling strengths and, in particular, allows us to consistently explore the extremely strong coupling regime for the first time. Our formalism elucidates difficulties of level truncations in the conventional frames from a general perspective, and  offers a systematic way to derive the faithful tight-binding Hamiltonians~\eqref{TBU}.    
While the emphasis was placed on the extremely strong coupling, our formalism is versatile enough to be applied to any coupling regimes, where standard/conventional descriptions can be inadequate.    
It would be interesting to apply the present formulation to identify the correct tight-binding models of more complex light-matter systems. In particular, it merits further study to elucidate role of the light-induced band flattening and narrowing in genuine many-body regimes.

\begin{acknowledgments}
We are grateful to Jerome Faist, Zongping Gong, and Giacomo Scalari  for fruitful discussions. Y.A. acknowledges support from the Japan Society for the Promotion of Science through Grant No.~JP19K23424. E.D. acknowledges support from Harvard-MIT CUA, AFOSR-MURI Photonic Quantum Matter (award FA95501610323), DARPA DRINQS program  (award D18AC00014), and the NSF EAGER-QAC-QSA award  2038011 ``Quantum Algorithms for Correlated Electron-Phonon System".
\end{acknowledgments}

\bibliography{reference}

\widetext
\pagebreak
\begin{center}
\textbf{\large Supplementary Materials}
\end{center}

\renewcommand{\theequation}{S\arabic{equation}}
\renewcommand{\thefigure}{S\arabic{figure}}
\renewcommand{\bibnumfmt}[1]{[S#1]}
\setcounter{equation}{0}
\setcounter{figure}{0}

\subsection{Polarization dependence of the effective length scale}
\begin{figure}[b]
\includegraphics[width=90mm]{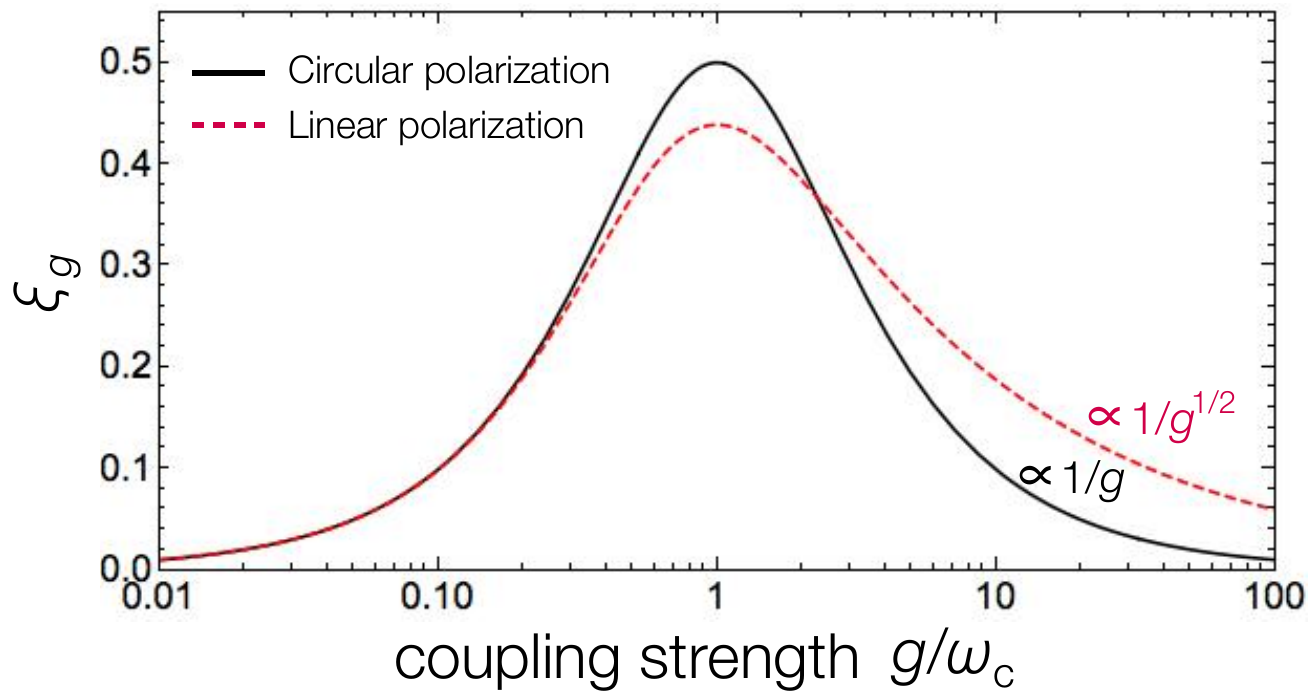} 
\caption{\label{fig1s}
Effective length scale $\xi_g$ for a circularly polarized light (black solid curve) and a linearly polarized light (red dashed curve) against the bare light-matter coupling  $g$. This length characterizes the effective interaction strength in the asymptotically decoupled frame. We set $\omega_c=\hbar=m=1$.
}
\end{figure}
We here mention that the effective length scale $\xi_g$, which characterizes the light-matter interaction strength in the asymptotically decoupled (AD) frame, in general depends on a polarization of an electromagnetic mode coupled to a many-body system. To demonstrate this, we consider a two-dimensional many-body system coupled to a circularly polarized electromagnetic mode as an illustrative example:
 \eqn{
\hat{\boldsymbol{A}}={\cal A}\left(\boldsymbol{e}\hat{a}+\boldsymbol{e}^{*}\hat{a}^{\dagger}\right),\;\;\boldsymbol{e}=\frac{1}{\sqrt{2}}\left[\begin{array}{c}
1\\
i
\end{array}\right].}
In this case, there are no terms that are proportional to $\hat{a}\hat{a}$ or $\hat{a}^\dagger\hat{a}^\dagger$ in the $\hat{\boldsymbol{A}}^2$ term;  this diamagnetic term simply renormalizes the photon  frequency without performing the Bogoliubov transformation. Thus, the resulting light-matter Hamiltonian in the Coulomb gauge is given by 
\eqn{
\hat{H}_{{\rm C}}=\int d\boldsymbol{x}\,\hat{\psi}_{\boldsymbol{x}}^{\dagger}\left(-\frac{\hbar^{2}\nabla^{2}}{2m}+V(\boldsymbol{x})\right)\hat{\psi}_{\boldsymbol{x}}-gx_{\omega_{c}}\int d\boldsymbol{x}\,\hat{\psi}_{\boldsymbol{x}}^{\dagger}(-i\hbar\nabla)\hat{\psi}_{\boldsymbol{x}}\cdot\left(\boldsymbol{e}\hat{a}+\boldsymbol{e}^{*}\hat{a}^{\dagger}\right)+\hbar\Omega\hat{a}^{\dagger}\hat{a},
}
where $g=q{\cal A}\sqrt{\omega_c/m\hbar}$ and we introduce 
\eqn{\label{circpolmic}
x_{\omega_{c}}=\sqrt{\frac{\hbar}{m\omega_{c}}},\;\;\Omega=\omega_{c}\left(1+\frac{Ng^{2}}{\omega_{c}^{2}}\right).
}
Note that the renormalized photon frequency depends on $g$ in a different way from the linearly polarized case discussed in the main text, for which $\Omega=\sqrt{\omega_c^2+2Ng^2}$. 

To asymptotically decouple the light and matter degrees of freedom, we can use a unitary transformation
\eqn{\label{unicirc}
\hat{U}=\exp\left[-i\xi_{g}\int d\boldsymbol{x}\,\hat{\psi}_{\boldsymbol{x}}^{\dagger}(-i\nabla)\hat{\psi}_{\boldsymbol{x}}\cdot\hat{\boldsymbol{\pi}}\right],\;\;\;\hat{\boldsymbol{\pi}}=i\left(\boldsymbol{e}^{*}\hat{a}^{\dagger}-\boldsymbol{e}\hat{a}\right),
}
where we define the effective length scale $\xi_g$ by
\eqn{
\xi_{g}=\frac{gx_{\omega_{c}}}{\Omega}=x_{\omega_{c}}\frac{g/\omega_{c}}{1+Ng^{2}/\omega_{c}^{2}}.
}
This length scale for a circularly polarized case asymptotically vanishes in the strong-coupling limit with the scaling $\propto 1/g$, which is faster than the linearly polarized case $\propto 1/g^{1/2}$ [see Fig.~\ref{fig1s}].
For the sake of completeness, we also show the full expression of the transformed Hamiltonian $\hat{H}_U=\hat{U}^\dagger\hat{H}_{\rm C}\hat{U}$ in the present case:
\eqn{
\hat{H}_{U}\!=\!\int d\boldsymbol{x}\,\hat{\psi}_{\boldsymbol{x}}^{\dagger}\left[-\frac{\hbar^{2}\nabla^{2}}{2m}+V\left(\boldsymbol{x}+\xi_{g}\hat{\boldsymbol{\pi}}\right)\right]\hat{\psi}_{\boldsymbol{x}}\!+\!\hbar\Omega\hat{a}^{\dagger}\hat{a}-\frac{\hbar^{2}g^{2}}{m\Omega^{2}}\left[\int d\boldsymbol{x}\,\hat{\psi}_{\boldsymbol{x}}^{\dagger}(-i\nabla)\hat{\psi}_{\boldsymbol{x}}\right]^{2}\!+\!\int d\boldsymbol{x}d\boldsymbol{x}'\,\frac{q^{2}\hat{\psi}_{\boldsymbol{x}}^{\dagger}\hat{\psi}_{\boldsymbol{x}'}^{\dagger}\hat{\psi}_{\boldsymbol{x}'}\hat{\psi}_{\boldsymbol{x}}}{4\pi\epsilon_{0}|\boldsymbol{x}-\boldsymbol{x}'|}.
}

\subsection{Enhancement of the effective mass}
We here demonstrate that the enhanced effective mass, which was discussed in the main text for a single-particle sector, in the case of a general $N$-particle system. For the sake of simplicity, we consider a 1D translationally invariant many-body system consisting of $N$ particles coupled to an electromagnetic mode, and write its Hamiltonian in the first-quantization form:
\eqn{
\hat{H}_{{\rm C}}=\sum_{j=1}^N\frac{(\hat{p}_{j}-q\hat{A})^{2}}{2m}+\sum_{j<j'}\frac{q^{2}}{4\pi\epsilon_{0}|x_{j}-x_{j'}|}+\hbar\omega_{c}\hat{a}^{\dagger}\hat{a}.
}
 Introducing the total momentum $\hat{P}=\frac{1}{\sqrt{N}}\sum_{j}\hat{p}_{j}$, we can decompose $\hat{H}_{{\rm C}}$ as
\eqn{
\hat{H}_{{\rm C}}=\hat{H}_{{\rm rel}}+\frac{\hat{P}^{2}}{2m}-\frac{q{\cal A}}{m}\sqrt{N}\hat{P}(\hat{a}+\hat{a}^{\dagger})+\frac{Nq^{2}{\cal A}^{2}}{2m}(\hat{a}+\hat{a}^{\dagger})^{2}+\hbar\omega_{c}\hat{a}^{\dagger}\hat{a},
}
where  $\hat{H}_{{\rm rel}}$ governs the relative motion of particles:
\eqn{
\hat{H}_{{\rm rel}}=\sum_{j=1}^N\frac{(\hat{p}_{j}-\hat{P}/\sqrt{N})^{2}}{2m}+\sum_{j<j'}\frac{q^{2}}{4\pi\epsilon_{0}|x_{j}-x_{j'}|}.
}
After performing the Bogoliubov transformation for the photon operator
and applying the decoupling unitary transformation in the same manner as done in the main text, we obtain
\eqn{
\hat{H}_{U} =\hat{H}_{{\rm rel}}+\left(1-\frac{2Ng^{2}}{\Omega^{2}}\right)\frac{\hat{P}^{2}}{2m}+\hbar\Omega\hat{a}^{\dagger}\hat{a}=\hat{H}_{{\rm rel}}+\frac{\hat{P}^{2}}{2m_{{\rm eff}}}+\hbar\Omega\hat{a}^{\dagger}\hat{a},
}
where we introduce the enhanced effective mass as
\eqn{
m_{{\rm eff}}=m(1+2Ng^{2}/\omega_{c}^{2}).
}
We note that this reduces to the effective mass discussed in the main text in the case of a single-particle sector $N=1$.

\subsection{Derivation of the tight-binding models}
We here provide technical details about the derivation of the tight-binding models in the AD frame and the Coulomb gauge. 
 We consider an electron that is subject to the periodic potential $V(x)=v[1+\cos(2\pi x/d)]$ and coupled to an electromagnetic mode. The total Hamiltonian in the AD frame is given by [cf. Eq.~(5) in the main text]
 \eqn{\label{hus}
 \hat{H}_{U}=\frac{\hat{p}^{2}}{2m_{{\rm eff}}}+v\left[1+\cos\left(Kx\right)\cos\left(K\xi_{g}\hat{\pi}\right)-\sin\left(Kx\right)\sin\left(K\xi_{g}\hat{\pi}\right)\right]+\hbar\Omega\hat{b}^{\dagger}\hat{b},
 }
 where $\hat{\pi}=i(\hat{b}^\dagger-\hat{b})$ and $K=\frac{2\pi}{d}$. 
 To derive the effective low-energy Hamiltonian, we consider the lowest-band Bloch wavefunctions for the single-particle Hamiltonian with the effective mass $m_{\rm eff}$:
 \eqn{
\left[\frac{\hat{p}^{2}}{2m_{{\rm eff}}}+V(x)\right]\psi_{k}=\epsilon_{k,g}\psi_{k},
 }
 where $\psi_{k}(x)=e^{ikx}u(x)$ with $u(x)$ satisfying $u(x)=u(x+d)$. Here, we note the $g$ dependence of the dispersion $\epsilon_{k,g}$, which comes through the effective mass $m_{\rm eff}=m[1+2(g/\omega_c)^2]$. 
 The corresponding Wannier function is
 \eqn{
w_{j}(x)=\int\frac{dk}{K}e^{-ikjd}\psi_{k}(x).
 }
 We now introduce a manifold of light-matter wavefunction spanned by product states consisting of these Wannier orbitals and an electromagnetic mode:
 \eqn{
|\Psi_{j}\rangle=\int dx\,w_{j}(x)|x\rangle\otimes|\psi_{{\rm photon}}\rangle,
 }
 where $|\psi_{\rm photon}\rangle$ represents a photon state. 
 When we consider the projection of $\hat{H}_U$ onto this manifold, the contribution from the term proportional to $\sin(Kx)$ in Eq.~\eqref{hus} vanishes. This is because the Hamiltonian $\hat{H}_U$ has the parity symmetry under $x\to -x$ and $\hat{\pi}\to-\hat{\pi}$ and the lowest states reside in the even parity sector. Since the lowest-band Wannier state $w(x)$ respects the even parity symmetry, a photon wavefunction must also be symmetric against $\hat{\pi}\to-\hat{\pi}$. This fact leads to $\langle\sin(K\xi_g\hat{\pi})\rangle=0$, where $\langle\cdots\rangle$ represents an expectation value with respect to a photon wavefunction with the even parity.  Thus, after performing the tight-binding approximation and taking into account the leading contributions, the projection results in the matrix elements
 \eqn{
\langle\Psi_{j}|\hat{H}_{U}|\Psi_{i}\rangle&=&\langle\Psi_{j}|\frac{\hat{p}^{2}}{2m_{{\rm eff}}}+V(x)+v\cos\left(Kx\right)\left[\cos\left(K\xi_{g}\hat{\pi}\right)-1\right]+\hbar\Omega\hat{b}^{\dagger}\hat{b}|\Psi_{i}\rangle\nonumber\\
	&\simeq &t_{g}\left(\delta_{i,j+1}+\delta_{i,j-1}\right)+\mu_{g}\delta_{i,j}+\left[t'_{g}\left(\delta_{i,j+1}+\delta_{i,j-1}\right)+\mu'_{g}\delta_{i,j}\right]\langle\hat{\delta}_{g}\rangle+\hbar\Omega\delta_{i,j}\langle\hat{b}^{\dagger}\hat{b}\rangle,
 }
 where we introduce the renormalized tight-binding parameters depending on $g$ as
 \eqn{t_{g}=\int\frac{dk}{K}\epsilon_{k,g}e^{ikd}\in\mathbb{R},\;\;\mu_{g}=\int\frac{dk}{K}\epsilon_{k,g},}
 \eqn{
t'_{g}=v\int dx\,w_{i-1}^{*}\cos(Kx)w_{i}\in\mathbb{R},\;\;\mu'_{g}=v\int dx\,w_{i}^{*}\cos(Kx)w_{i},
 }
 and the operator describing the electromagnetically induced fluctuation by
 \eqn{
\hat{\delta}_g=\cos\left(K\xi_{g}\hat{\pi}\right)-1.
 } 
 After transforming to the second quantization notation, we obtain the tight-binding Hamiltonian in the AD frame, which provides Eq.~(9) in the main text
 \eqn{\label{tbsup}
\hat{H}_{U}^{{\rm TB}}=\left(t_{g}+t'_{g}\hat{\delta}_g\right)\sum_{i}\left(\hat{c}_{i}^{\dagger}\hat{c}_{i+1}+{\rm h.c.}\right)+\left(\mu_{g}+\mu'_{g}\hat{\delta}_g\right)\sum_{i}\hat{c}_{i}^{\dagger}\hat{c}_{i}+\hbar\Omega\hat{b}^{\dagger}\hat{b},
 }
 where the annihilation operator should be understood in terms of the expansion $\hat{\psi}_{x}=\sum_{j}w_{j}(x)\hat{c}_{j}$.
 We remark that, while this tight-binding description is valid when low-energy equilibrium properties are of interest, one may have to include further correction terms for analyzing nonequilibrium dynamics.  For instance, the contribution from the $\sin(Kx)$ term in Eq.~\eqref{hus} can be relevant when excitations to higher bands are nonnegligible. 
 
\begin{figure}[t]
\includegraphics[width=140mm]{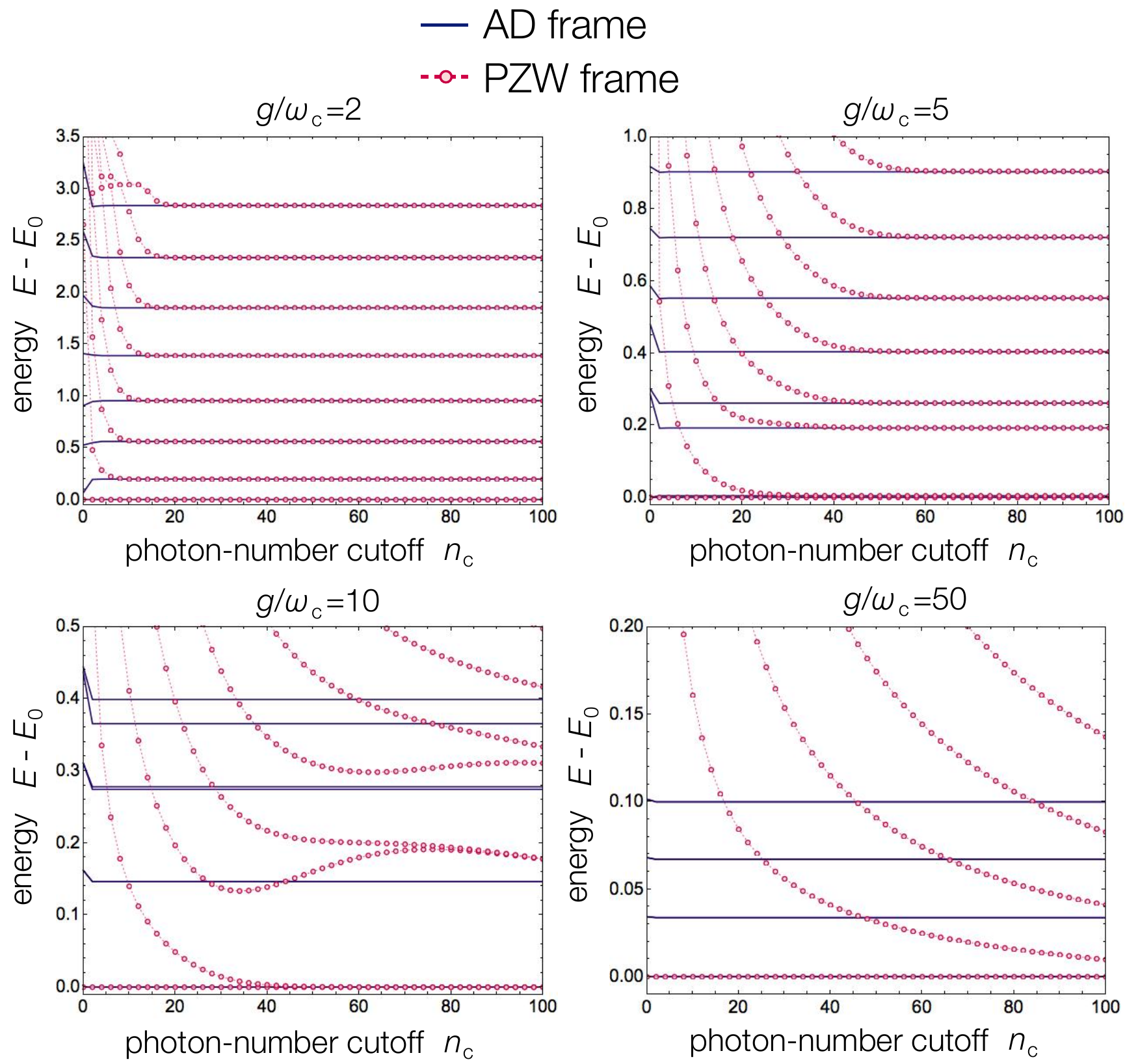} 
\caption{\label{fig2s}
Comparisons of convergence of low-energy spectra in the AD frame (blue solid curves) and the PZW frame (red dotted curves) with respect to the photon-number cutoff $n_c$ at different coupling strengths $g$. The AD-frame energies efficiently converge at low $n_c$, while the results in the PZW frame require an increasingly large cutoff at stronger $g$, and do not converge for $g/\omega_c\gtrsim 10$ at least in the plotted scale. We set $\omega_c=\sqrt{\hbar/m\omega_c}=1$ and choose the parameters $\lambda\!=\!3$, $\mu\!=\!3.85$.   
}
\end{figure}

  We here note that the calculation of matrix elements of the shifted potential term $V(x+\xi_g\hat{\pi})$ has been separated into light and matter parts. This simplification has its origin in the translationally invariance of the potential $V(x)$, and can be transferred to a general periodic potential.
  While one needs to work with the operator-valued term such as $\cos(K\xi_g\hat{\pi})$, in practice, it can still efficiently be calculated since it usually suffices to set a photon-number cutoff to at most $\sim100$ in the AD frame, for which $\hat{\pi}$ is just a matrix with a small dimension [cf. Fig.~\ref{fig2s}]. This is the reason why we did not need to rely on the approximative form (Eq.~(10) in the main text) obtained by the expansion in $\xi_{g}\hat{\pi}$, but only on the tight-binding approximation, resulting in Eq.~\eqref{tbsup}.
   
 Meanwhile, when we are interested in a nonperiodic potential, the simplification at a large coupling can be made possible by performing the Taylor expansion and truncating at a finite order as discussed in the main text.  Said differently, the calculation of the shifted potential can be challenging if the potential is nonperiodic and singular such that the truncation cannot be well-justified and the coupling strength lies in the intermediate regime $g/\omega_c\sim 1$ for which the effective length $\xi_g$ is rather large [cf. Fig.~1 in the main text].

 For the sake of comparison, we next explain the construction of the tight-binding Hamiltonian in the Coulomb gauge. We start from the Coulomb-gauge Hamiltonian
 \eqn{
\hat{H}_{{\rm C}}=\frac{\hat{p}^{2}}{2m}+V(x)-gx_{\Omega}\hat{p}(\hat{b}+\hat{b}^{\dagger})+\hbar\Omega\hat{b}^{\dagger}\hat{b}.
 }
 Similar to the above discussion, we consider the lowest-band Wannier states for the single-particle Hamiltonian with the {\it bare} mass $m$:
 \eqn{
\left[\frac{\hat{p}^{2}}{2m}+V\left(x\right)\right]\tilde{\psi}_{k}=\tilde{\epsilon}_{k}\tilde{\psi}_{k},\;\;\tilde{w}_{j}(x)=\int\frac{dk}{K}e^{-ikjd}\tilde{\psi}_{k}(x),
 }
 and consider a manifold spanned by the following light-matter states
  \eqn{
 |\tilde{\Psi}_{j}\rangle=\int dx\,\tilde{w}_{j}(x)|x\rangle\otimes|\psi_{{\rm photon}}\rangle.
  }
  We note that the dispersion $\tilde{\epsilon}_k$ is independent of $g$ as we here consider the bare mass $m$.
The projection of $\hat{H}_{\rm C}$ onto this manifold results in the matrix elements 
\eqn{
\langle\tilde{\Psi}_{j}|\hat{H}_{{\rm C}}|\tilde{\Psi}_{i}\rangle	&=&\langle\tilde{\Psi}_{j}|\frac{\hat{p}^{2}}{2m}+V(x)-gx_{\Omega}\hat{p}(\hat{b}+\hat{b}^{\dagger})+\hbar\Omega\hat{b}^{\dagger}\hat{b}|\tilde{\Psi}_{i}\rangle\nonumber\\
	&\simeq&\tilde{t}\left(\delta_{i,j+1}+\delta_{i,j-1}\right)+\tilde{\mu}\delta_{i,j}-\left(\tilde{\lambda}_{g}\delta_{i,j+1}+\tilde{\lambda}_{g}^{*}\delta_{i,j-1}\right)\langle\hat{b}+\hat{b}^{\dagger}\rangle+\hbar\Omega\delta_{i,j}\langle\hat{b}^{\dagger}\hat{b}\rangle,
}
where $\langle\cdots\rangle$ represents an expectation value with respect to an arbitrary photon state and the tight-binding parameters are defined by
\eqn{
\tilde{t}=\int\frac{dk}{K}\tilde{\epsilon}_{k}e^{ikd}\in\mathbb{R},\;\;\tilde{\mu}=\int\frac{dk}{K}\tilde{\epsilon}_{k},\;\;\tilde{\lambda}_{g}=\hbar gx_{\Omega}\int dx\,\tilde{w}_{i-1}^{*}(-i\partial_{x})\tilde{w}_{i}\in i\mathbb{R}.
}
We again emphasize that, in contrast to the AD-frame case above, the tight-binding parameters are defined in terms of the single-particle states with the bare mass $m$; thus, in particular, $\tilde{t},\tilde{\mu}$ are independent of the light-matter coupling $g$. 

In the second quantization notation, the tight-binding Hamiltonian can be written as
\eqn{
\hat{H}_{{\rm C}}^{{\rm TB}}=\sum_{i}\left(\left[\tilde{t}-\tilde{\lambda}_{g}\left(\hat{b}+\hat{b}^{\dagger}\right)\right]\hat{\tilde{c}}_{i}^{\dagger}\hat{\tilde{c}}_{i+1}+{\rm h.c.}\right)+\tilde{\mu}\sum_{i}\hat{\tilde{c}}_{i}^{\dagger}\hat{\tilde{c}}_{i}+\hbar\Omega\hat{b}^{\dagger}\hat{b},
}
where the annihilation operator is defined in terms of the Wannier function with the bare mass, $\hat{\psi}_{x}=\sum_{j}\tilde{w}_{j}(x)\hat{\tilde{c}}_{j}$. 
Its eigenspectrum can analytically be given by
\eqn{\label{dispC}
\tilde{\epsilon}_{k,n,{\rm C}}^{{\rm TB}}=2\tilde{t}\cos\left(kd\right)+\tilde{\mu}-\frac{4\tilde{\lambda}_{g}^{2}}{\Omega}\sin^{2}\left(kd\right)+\hbar\Omega\,n,
}
where $n=0,1,2\ldots$ The results plotted in Fig.~2 in the main text correspond to the $n=0$ sector of this dispersion.  It is evident from Eq.~\eqref{dispC} that the tight-binding spectrum in the Coulomb gauge is completely independent of $g$ at $k=0,\pm \pi/d$, which clearly indicates difficulties of level truncations in the Coulomb gauge.

Finally, we remark that the one-dimensional tight-binding model acquires additional contributions in the case of the circularly polarized light. To see this, it is sufficient to consider the following single-particle continuum model [see Eq.~\eqref{circpolmic} for the definitions of the microscopic parameters]:
\eqn{
\hat{H}_{{\rm C}}=\frac{\hat{\boldsymbol{p}}^2}{2m}+V(x)+\frac{m\Omega_y^2 y^2}{2}-gx_{\omega_{c}}\hat{\boldsymbol{p}}\cdot\left(\boldsymbol{e}\hat{a}+\boldsymbol{e}^{*}\hat{a}^{\dagger}\right)+\hbar\Omega\hat{a}^{\dagger}\hat{a},
}
where $\boldsymbol{e}=\frac{1}{\sqrt{2}}[1,i]^{\rm T}$ is the polarization vector, $\Omega=\omega_c(1+g^2/\omega_c^2)$, and the electron is tightly localized in the transverse $y$ direction via the potential $m\Omega_y^2 y^2/2$, while it is subject to the periodic potential $V(x)$ in the $x$ direction. Using the unitary transformation~\eqref{unicirc}, we obtain 
\eqn{\label{circtbm}
\hat{H}_{U}\!=\!\frac{\hat{\boldsymbol{p}}^2}{2m_{\rm eff}}+V\left(x+i\xi_g(\hat{a}^\dagger-\hat{a})/\sqrt{2}\right)+\frac{m\Omega_y^2 \left[y+\xi_g(\hat{a}+\hat{a}^\dagger)/\sqrt{2}\right]^2}{2}+\hbar\Omega\hat{a}^{\dagger}\hat{a},
}
where $m_{\rm eff}=m[1+(g/\omega_c)^2]$ and $\xi_g=gx_{\omega_c}/\Omega=x_{\omega_c}g/(\omega_c+g^2/\omega_c)$. It is now clear that, even when we are interested in 1D electron dynamics in the $x$ direction and aim to derive the tight-binding model along this direction, we must in general take into account the contribution from the third term in the RHS of Eq.~\eqref{circtbm} that arises from the coupling between the transverse motion and the circularly polarized light. Nevertheless, the simple 1D description along the $x$ direction analogous to Eq.~(5)  in the main text (i.e., neglecting the orbital motion along the $y$ direction) can still be recovered when (i) $\Omega_y$ is large so that motional excitation in the $y$ direction is suppressed and (ii) the coupling $g$ is large in the sense that $\Omega_y\ll g^2/\omega_c$ (which means  $m\Omega_y^2\xi_g^2\ll\hbar\Omega$) in such a way that the light-matter coupled term can be neglected compared to the dressed photon term. We note that the condition (ii) can in principle be satisfied for any finite $\Omega_y$ provided that $g$ is sufficiently large.

\subsection{Photon-number cutoff dependence of the low-energy spectra}
We here briefly mention the photon-number cutoff dependence of the low-energy spectra in different frames. We compare the spectra for the double-well potential $V=-\lambda x^2/2+\mu x^4/4$ in the AD frame $\hat{H}_U=\hat{U}^\dagger\hat{H}_{\rm C}\hat{U}$ with $\hat{U}=\exp(-i\xi_g\hat{p}\hat{\pi}/\hbar)$ and the Power-Zienau-Woolley (PZW) frame $\hat{H}_{{\rm PZW}}=\hat{U}_{{\rm PZW}}^{\dagger}\hat{H}_{{\rm C}}\hat{U}_{{\rm PZW}}$ with $\hat{U}_{\rm PZW}=\exp(iqx\hat{A}/\hbar)$:
\eqn{
\hat{H}_{U}&=&\frac{\hat{p}^{2}}{2m_{{\rm eff}}}+V\left(x+\xi_{g}\hat{\pi}\right)+\hbar\Omega\hat{b}^{\dagger}\hat{b},\label{huss}\\
\hat{H}_{{\rm PZW}}&=&\frac{\hat{p}^{2}}{2m}+V(x)+mg^{2}x^{2}+ig\sqrt{m\hbar\omega_{c}}x(\hat{a}^\dagger-\hat{a})+\hbar\omega_{c}\hat{a}^{\dagger}\hat{a}.
}
In Fig.~3 in the main text, we show the results at the large photon-number cutoff $n_c=100$, and demonstrate that the PZW frame fails to capture the key features in the extremely strong coupling (ESC) regime, such as the level degeneracy and narrowing. This is caused by the slower convergence of the PZW results  at larger coupling $g$ with respect to the photon-number cutoff $n_c$. To see this explicitly, we plot the low-lying energies (subtracted by the lowest eigenvalue) in different frames in Fig.~\ref{fig2s}. While the results in the AD frame efficiently converge already for low cutoff $n_c\sim 5\!-\!10$ at any coupling strength $g$, the convergence in the PZW frame becomes worse as $g$ is increased. In particular, in the ESC regime (roughly corresponding to $g/\omega_c\gtrsim 10$), the PZW results typically fail to converge within a tractable value of the photon-number cutoff. This difficulty stems from the rapid increase of the mean-photon number in an energy eigenstate due to large entanglement among high-lying levels present in the PZW frame.

\subsection{Derivation of the multimode generalization of the unitary transformation}
We provide details about the derivation of the multimode generalization of our formalism presented in the main text. 
We start from the light-matter Hamiltonian including multiple spatially varying electromagnetic modes in the Coulomb gauge:
\eqn{
\hat{H}_{{\rm C}}=\frac{\hat{\boldsymbol{p}}^{2}}{2m}+V(\boldsymbol{x})-\frac{q}{2m}\left(\hat{\boldsymbol{p}}\cdot\hat{\boldsymbol{A}}(\boldsymbol{x})+{\rm h.c.}\right)+\frac{q^{2}\hat{\boldsymbol{A}}^{2}(\boldsymbol{x})}{2m}+\sum_{\boldsymbol{k}\lambda}\hbar\omega_{\boldsymbol{k}}\hat{a}_{\boldsymbol{k}\lambda}^{\dagger}\hat{a}_{\boldsymbol{k}\lambda},
}
where we consider the vector potential expanded by plane waves
\eqn{
\hat{\boldsymbol{A}}(\boldsymbol{x})=\sum_{\boldsymbol{k}\lambda}\boldsymbol{\epsilon}_{\boldsymbol{k}\lambda}{\cal A}_{\boldsymbol{k}}\left(\hat{a}_{\boldsymbol{k}\lambda}e^{i\boldsymbol{k}\cdot\boldsymbol{x}}+{\rm h.c.}\right),\;\;\boldsymbol{k}\cdot\boldsymbol{\epsilon}_{\boldsymbol{k}\lambda}=0,\;\;\boldsymbol{\epsilon}_{\boldsymbol{k}\lambda}\cdot\boldsymbol{\epsilon}_{\boldsymbol{k}\nu}=\delta_{\lambda\nu}
}
with $\lambda$ denoting polarization.
To generalize the asymptotically decoupling unitary transformation~$\hat{U}$ to this multimode case, we first introduce the field operators
\eqn{
\hat{X}_{\boldsymbol{k}\lambda}(\boldsymbol{x})\equiv\sqrt{\frac{\hbar}{2\omega_{\boldsymbol{k}}}}\left(\hat{a}_{\boldsymbol{k}\lambda}e^{i\boldsymbol{k}\cdot\boldsymbol{x}}+\hat{a}_{\boldsymbol{k}\lambda}^{\dagger}e^{-i\boldsymbol{k}\cdot\boldsymbol{x}}\right),\;\hat{P}_{\boldsymbol{k}\lambda}(\boldsymbol{x})\equiv\sqrt{\frac{\hbar\omega_{\boldsymbol{k}}}{2}}i\left(\hat{a}_{\boldsymbol{k}\lambda}^{\dagger}e^{-i\boldsymbol{k}\cdot\boldsymbol{x}}-\hat{a}_{\boldsymbol{k}\lambda}e^{i\boldsymbol{k}\cdot\boldsymbol{x}}\right),
}
and define the coupling as
\eqn{
g_{\boldsymbol{k}}=q{\cal A}_{\boldsymbol{k}}\sqrt{\frac{\omega_{\boldsymbol{k}}}{m\hbar}}.
}
We then rewrite the quadratic photon part of the Hamiltonian (aside the constant) as 
\eqn{
\frac{q^{2}\hat{\boldsymbol{A}}^{2}(\boldsymbol{x})}{2m}+\sum_{\boldsymbol{k}\lambda}\hbar\omega_{\boldsymbol{k}}\hat{a}_{\boldsymbol{k}\lambda}^{\dagger}\hat{a}_{\boldsymbol{k}\lambda}&=&\sum_{\boldsymbol{k}\lambda}\frac{\hat{P}_{\boldsymbol{k}\lambda}^{2}(\boldsymbol{x})}{2}+\frac{1}{2}\sum_{\boldsymbol{k}\lambda\boldsymbol{k}'\lambda'}\left(\delta_{\boldsymbol{k}\lambda,\boldsymbol{k}'\lambda'}\omega_{\boldsymbol{k}}^{2}+2g_{\boldsymbol{k}}g_{\boldsymbol{k}'}\boldsymbol{\epsilon}_{\boldsymbol{k}\lambda}\cdot\boldsymbol{\epsilon}_{\boldsymbol{k}'\lambda'}\right)\hat{X}_{\boldsymbol{k}\lambda}(\boldsymbol{x})\hat{X}_{\boldsymbol{k}'\lambda'}(\boldsymbol{x})\nonumber\\
&=&\frac{1}{2}\sum_{\alpha}\left(\hat{P}_{\alpha}^{2}(\boldsymbol{x})+\Omega_{\alpha}^{2}\hat{X}_{\alpha}^{2}(\boldsymbol{x})\right),
}
where we define the diagonalized basis labeled by $\alpha$ via
\eqn{
\hat{X}_{\boldsymbol{k}\lambda}(\boldsymbol{x})=\sum_{\alpha}O_{\boldsymbol{k}\lambda,\alpha}\hat{X}_{\alpha}(\boldsymbol{x})
}
with $O_{\boldsymbol{k}\lambda,\alpha}$ being an orthogonal matrix. 

We next introduce the $\boldsymbol{x}$-dependent annihilation operators via
\eqn{
\hat{b}_{\alpha}(\boldsymbol{x})\equiv\sqrt{\frac{\Omega_{\alpha}}{2\hbar}}\hat{X}_{\alpha}(\boldsymbol{x})+\frac{i}{\sqrt{2\hbar\Omega_{\alpha}}}\hat{P}_{\alpha}(\boldsymbol{x}),
}
and also define the vector-valued variables labeled by $\alpha$ as
\eqn{
\boldsymbol{\zeta}_{\alpha}=x_{\Omega_{\alpha}}\sum_{\boldsymbol{k}\lambda}\boldsymbol{\epsilon}_{\boldsymbol{k}\lambda}g_{\boldsymbol{k}}O_{\boldsymbol{k}\lambda,\alpha},
}
where $x_{\Omega_{\alpha}}=\sqrt{\frac{\hbar}{m\Omega_{\alpha}}}$. 
We now introduce the unitary transformation in the multimode case by
\eqn{
\hat{U}=\exp\left[-i\frac{\hat{\boldsymbol{p}}}{\hbar}\cdot\sum_{\alpha}\boldsymbol{\xi}_{\alpha}\hat{\pi}_{\alpha}(\boldsymbol{x})\right],\;\;\hat{\pi}_{\alpha}(\boldsymbol{x})=i\left(\hat{b}_{\alpha}^{\dagger}(\boldsymbol{x})-\hat{b}_{\alpha}(\boldsymbol{x})\right),\;\;\boldsymbol{\xi}_{\alpha}=\frac{\boldsymbol{\zeta}_{\alpha}}{\Omega_{\alpha}}.
}
We note that, since the electromagnetic modes now explicitly depend on the position $\boldsymbol{x}$, they do not commute with the momentum operator $\hat{\boldsymbol{p}}$ in the transformation $\hat{U}$, and thus, the transformed Hamiltonian in general acquires additional contributions compared to the simple expression obtained in the single-mode case [cf. Eq.~\eqref{huss}]. 
Nevertheless, significant simplification can occur when the field variation is small compared with the effective length scale:
\eqn{\label{cond}
k|\boldsymbol{\xi}|\ll 1\;\;{\rm for}\;\;|\nabla\hat{b}|\sim k\hat{b}.
}
We emphasize that this condition is independent of system size and thus much less restrictive than the standard dipole approximation. In particular, Eq.~\eqref{cond} can, in principle, be attained for any $k$ if the coupling $g$ is taken to be sufficiently strong such that the effective length scale $|\boldsymbol{\xi}|$ is short enough to satisfy this condition. 
Under this condition, the derivative terms of the field operators $\hat{b}(\boldsymbol{x})$, $\hat{\pi}(\boldsymbol{x})$ can be neglected, resulting in the simple transformed Hamiltonian:
\eqn{
\hat{H}_{U}=\hat{U}^{\dagger}\hat{H}_{C}\hat{U}\simeq\frac{\hat{\boldsymbol{p}}^{2}}{2m}-\sum_{\alpha}\frac{\left(\hat{\boldsymbol{p}}\cdot\boldsymbol{\zeta}_{\alpha}\right)^{2}}{\hbar\Omega_{\alpha}}+V\left(\boldsymbol{x}+\sum_{\alpha}\boldsymbol{\xi}_{\alpha}\hat{\pi}_{\alpha}\left(\boldsymbol{x}\right)\right)+\sum_{\alpha}\hbar\Omega_{\alpha}\hat{b}_{\alpha}^{\dagger}(\boldsymbol{x})\hat{b}_{\alpha}(\boldsymbol{x}),
} 
which provides Eq.~(12) in the main text.

\end{document}